# Bayesian EWMA and CUSUM Control Charts Under Different Loss Functions

**Chelsea L. Mitchell**[a], **Abdel-Salam G. Abdel-Salam**[b] **and D'Arcy Mays**[a]

[a]*Department of Statistical Science and Operations Research*

*College of Humanities and Sciences*

*Virginia Commonwealth University, Richmond, VA*

[b]*Department of Mathematics, Statistics and Physics, Qatar University, Qatar*

## 1. Introduction

In statistical process monitoring (SPM), control charts are a vital visual tool used to monitor a process and alert of any discrepancies. Many control charts have been developed to encompass the different processes to monitor, mainly falling into two categories: memory-less and memory-based. A memory-less chart does not take into consideration previous observations and are best used to detect large/sudden changes. Memory-based charts compute the current statistic using the previous' and are ideal for detecting small/gradual changes. The Shewhart control chart, first outlined in Shewhart (1926), is a frequently used memory-less chart while the cumulative sum (CUSUM) (Page, 1954) and exponentially weighted moving average (EWMA) (Roberts, 1959) charts are well-known memory-based charts.

In Riaz et al. (2017), the EWMA control chart is considered under a Bayesian approach. A comparison analysis is conducted on control limits found using the classical approach and the

Bayesian approach under a symmetric loss function and 2 asymmetric loss functions: squared error loss function (SELF), precautionary loss function (PLF), and Linex loss function (LLF). A normal conjugate prior is used on data that is normally distributed throughout the paper. For each of the loss functions, a sensitivity analysis is done for the choice of hyper-parameter and its effect on the performance of the Bayesian EWMA chart. The measurement tools used for comparative purposes are the average run length (ARL) and standard deviation run length (SDRL). Our work will conduct analysis of a Bayesian EWMA and a Bayesian CUSUM chart via ARL, SDRL, average time to signal (ATS), and standard deviation of time to signal (SDTS) using different likelihood and prior distributions under the same loss functions.

We first conduct analysis of the Bayesian CUSUM chart under the Normal conjugate case and compare results to that obtained using the Bayesian EWMA. Then, we complete an analysis of the respective control charts under a Poisson likelihood and Gamma prior. Lastly, we use the Poisson likelihood with an Exponential prior for analysis to show generality of the use of the control charts. Our objective is to assess the choice of loss function in relation to distribution and to deduce generality of the Bayesian EWMA and Bayesian CUSUM control charts.

## 2. Bayesian Inference

In 1763 an article entitled "An Essay Towards Solving a Problem in the Doctrine of Chances" was published posthumously by English statistician Thomas Bayes, introducing Bayes Theorem. The theorem outlines the probability of an event based on prior knowledge related to the event and has led to a branch of statistics which makes inferences structured from this concept. Bayes theorem combines the likelihood function, which is determined by the data, with the expert chosen prior distribution to determine what the posterior distribution will be. Once the posterior distribution is found, it can be combined with the likelihood of future $y$ data, $p(y|\mathbf{X})$, and integrated with respect to $\theta$ to obtain the posterior predictive distribution (eq. 2).

$$p(\theta|\mathbf{X}) = \frac{\mathbf{p}(\mathbf{X}|\theta)\mathbf{p}(\theta)}{\mathbf{p}(\mathbf{X})} \qquad (1)$$

|   | Definition |
|---|---|
| $\theta$ | Parameter(s) of interest |
| $\mathbf{X}$ | Data |
| $P(\theta)$ | Prior distribution |
| $P(\mathbf{X})$ | Marginal distribution data |
| $P(\mathbf{X}|\theta)$ | Likelihood function |

Table 1: Bayesian Inference Definitions

$$p(y|\mathbf{X}) = \int p(\mathbf{y}|\theta)\mathbf{p}(\theta|\mathbf{X})\mathbf{d}\theta \qquad (2)$$

where, $p(\mathbf{y}|\theta)$ is the likelihood function for the future data

$$\hat{\theta}^* = \min_{\hat{\theta}} E_\theta[L(\hat{\theta},\theta)] \qquad (3)$$

where, $\hat{\theta}^*$ is the estimator which minimizes the expected loss

## *2.1. Loss Functions*

A loss function is represented as $L(\theta,\hat{\theta})$ and is a function of the parameter of interest, $\theta$, and an estimate of the parameter, $\hat{\theta}$. Loss functions measure how bad the current parameter estimate is, and typically the larger the loss the worse the parameter estimate. In decision theory it takes on a slightly different role, which is to attain the best estimator of the parameter. In this study we consider three different loss functions: Squared Error, Precautionary, and Linex. We construct a Bayesian EWMA and Bayesian CUSUM under the different loss functions.

The Squared Error Loss Function (SELF) is a commonly used loss function because its simplicity. It is considered symmetric because it assigns equal weight to both positive and negative error and defined as $L(\theta,\hat{\theta}) = (\theta - \hat{\theta})^2$ with a Bayes estimator of $\hat{\theta}^* = E[\theta|x]$. The Precautionary Loss Function (PLF) is an asymmetric loss function as it weights the error in the positive direc-tion differently than in the negative direction. It was introduced in Norstrom (1996) in order to prevent overestimation of the parameter and typically produces better estimators for low failure rate problems. Its form is $L(\theta,\hat{\theta}) = \frac{(\theta-\hat{\theta})^2}{\hat{\theta}}$ with its Bayes estimator given as $\hat{\theta}^* = \sqrt{E[\theta^2|x]}$. The last loss function that we consider here is another asymmetric function introduced in Varian (1975) as the Linex Loss Function (LLF). It is asymmetric as it signifies overestimation rather than underestimation, and is often used with estimating the location parameter. It is defined as $L(\theta,\hat{\theta}) = (e^{c(\hat{\theta}-\theta)} - c(\hat{\theta}-\theta) - 1)$ with a Bayes estimator of $\hat{\theta}^* = -\frac{1}{c}lnE[e^{-c\theta}]$ (Zellner, 1986), where $c$ is a constant such that if $c > 0$ overestimation has more significance than underestimation and if $c \to 0$ it approaches symmetry.

## 3. Bayesian EWMA and Bayesian CUSUM Control Chart Schematics

The control limits for the Bayesian EWMA and Bayesian CUSUM control charts follow the same form as the standard EWMA and CUSUM charts respectively. The mean and standard deviation used for the chart limits are based on the mean and standard deviation obtained using the Bayes estimator posterior predictive distribution under each loss function. In equations 4 and 5, $\mu_{LF}$ and $\sigma_{\bar{Y}}$ are the mean and standard deviation from the Bayes estimator posterior predictive distribution respective to the loss function used and $\tau$ is a user-defined constant determining the amount of memory. The plotting statistics also follow the standard form for the EWMA and CUSUM control charts. In these equations $(\bar{y}|x)$ is the Bayes estimator for the set of predicted data.

EWMA Posterior Predictive Control Limits:

$$UCL/LCL = \mu_{LF} \pm L\sigma_{\bar{Y}}\sqrt{\frac{\tau}{2-\tau}}$$

$$CL = \mu_{LF}$$

$$z_i = \tau(\bar{y}|x) + (1-\tau)z_{i-1}$$

(4)

CUSUM Posterior Predictive Control Limits:

$$UCL/LCL = \pm h * \sigma_{\bar{Y}}$$

$$CL = \mu_{LF} \tag{5}$$

$$c_i = [(\bar{y}|x) - \mu_{LF}] + c_{i-1}$$

# 4. Bayesian Framework Under Different Loss Functions and Distributions

## 4.1. Normal Conjugate Prior

The calculation of the variances for the posterior, posterior predictive, and Bayes estimator of the posterior predictive distributions remain the same under each loss function. In table 2 $\sigma_0^2$ is the prior variance and $\sigma^2$ is the known variance of the likelihood distribution.

| Distribution | Variance |
| --- | --- |
| Posterior | $\sigma_n^2 = \frac{\sigma^2 \sigma_0^2}{\sigma^2 + n\sigma_0^2}$ |
| Posterior Predictive | $\sigma_{ppd}^2 = \sigma^2 + \sigma_n^2$ |
| Bayes Estimator of Posterior Predictive | $\sigma_{\bar{Y}}^2 = \frac{\sigma^2}{n} + \sigma_n^2$ |

Table 2: Variances Based on Distribution

### 4.1.1. Squared Error Loss Function

Posterior Distribution:

In equations 6, 7, and 8 $\mu_{SELF}$ is the mean derived from the Bayes estimator equation for the squared error loss function (Riaz et al., 2017).

$$\mu|\mathbf{X} \sim \mathbf{N}(\mu_{\mathbf{SELF}}, \sigma_\mathbf{n}^2) \text{ where,}$$
$$\mu_{SELF} = \frac{n\bar{x}\sigma_0^2 + \sigma^2\mu_0}{\sigma^2 + n\sigma_0^2} \tag{6}$$

Posterior Predictive Distribution:

$$\mathbf{Y}|\mathbf{X} \sim \mathbf{N}(\mu_{\mathbf{SELF}}, \sigma_{\mathbf{ppd}}^2) \tag{7}$$

Bayes Estimator for Posterior Predictive:

$$\bar{\mathbf{Y}}|\mathbf{X} \sim N(\mu_{SELF}, \sigma_{\bar{Y}}^2) \tag{8}$$

### 4.1.2. Precautionary Loss Function

Posterior Distribution:

In equations 9, 10, and 11 $\mu_{PLF}$ is the mean derived from the Bayes estimator equation for the squared error loss function (Riaz et al., 2017).

$$\mu|\mathbf{X} \sim \mathbf{N}(\mu_{\mathbf{PLF}}, \sigma_\mathbf{n}^2) \text{ where,}$$
$$\mu_{PLF} = \sqrt{\frac{(\sigma^4 + \sigma^2\sigma_0^2(n+1))(\sigma^2 + n\sigma_0^2) + (n\bar{x}\sigma_0^2 + \sigma^2\mu_0)^2}{(\sigma^2 + n\sigma_0^2)^2}} \tag{9}$$

Posterior Predictive Distribution:

$$\mathbf{Y}|\mathbf{X} \sim \mathbf{N}(\mu_{\mathbf{PLF}}, \sigma_{\mathbf{ppd}}^2) \tag{10}$$

Bayes Estimator for Posterior Predictive:

$$\bar{\mathbf{Y}}|\mathbf{X} \sim \mathbf{N}(\mu_{\mathbf{PLF}}, \sigma_{\bar{\mathbf{Y}}}^2) \tag{11}$$

### *4.1.3. Linex Loss Function*

In equations 12, 13, and 14 $\mu_{LLF}$ is the mean derived from the Bayes estimator equation for the squared error loss function (Riaz et al., 2017).

Posterior Distribution:

$$\mu|\mathbf{X} \sim \mathbf{N}(\mu_{\mathbf{LLF}}, \sigma_{\mathbf{n}}^{\mathbf{2}}) \text{ where,}$$
$$\mu_{LLF} = \frac{n\bar{x}\sigma_0^2 + \sigma^2\mu_0}{\sigma^2 + n\sigma_0^2} - \frac{c}{2}\left(\sigma^2 + \frac{\sigma^2\sigma_0^2}{\sigma^2 + n\sigma_0^2}\right) \quad (12)$$

Posterior Predictive Distribution:

$$\mathbf{Y}|\mathbf{X} \sim \mathbf{N}(\mu_{\mathbf{LLF}}, \sigma_{\mathbf{ppd}}^{\mathbf{2}}) \quad (13)$$

Bayes Estimator for Posterior Predictive:

$$\bar{\mathbf{Y}}|\mathbf{X} \sim \mathbf{N}(\mu_{\mathbf{LLF}}, \sigma_{\bar{\mathbf{Y}}}^{\mathbf{2}}) \quad (14)$$

## *4.2. Poisson Conjugate Prior*

The Gamma distribution is the known conjugate when used with the Poisson likelihood function. The posterior distribution under each of the loss functions will be Gamma with shape parameter $n\bar{x} + \alpha$ and inverse scale parameter $n + \beta$. The posterior predictive and Bayes estimator of the posterior predictive distributions are derived to be Negative Binomial (see 28) also with shape parameter $n\bar{x} + \alpha$ and inverse scale parameter $n + \beta$. Similarly as stated in section 4.1 the variances for each of the distributions in table 3 will remain the same under each loss function. In the table, $\alpha$ and $\beta$ are the shape and inverse scale parameters respectively. Prior information of the mean and variance are used to solve for the $\alpha$ and $\beta$ values. The mean for each of the distributions is what differs based on the Bayes' estimator equations for the given loss functions.

| Distribution | Variance |
|---|---|
| Posterior | $\sigma^2_{PD} = \frac{n\bar{x}+\alpha}{(n+\beta)^2}$ |
| Posterior Predictive | $\sigma^2_{PPD} = \frac{n\bar{x}+\alpha}{(n+\beta)^2}(n+\beta+1)$ |
| Bayes Estimator of Posterior Predictive | $\sigma^2_{\bar{Y}} = \frac{n\bar{x}+\alpha}{n(n+\beta)^2}(n+\beta+1)$ |

Table 3: Variances Based on Distribution

### 4.2.1. Squared Error Loss Function

In equations 15, 16, and 17 the mean, $\mu_{SELF,PG}$ (see 30), is the Bayes' estimator found by using the appropriate equation found in section 2.1.

Posterior Distribution:

$$\lambda|\mathbf{X} \sim \mathbf{Gamma}(\mathbf{n}\bar{x} + \alpha, n + \beta) \text{ where,}$$
$$\mu_{SELF,PG} = \frac{n\bar{x} + \alpha}{n + \beta} \tag{15}$$

Posterior Predictive Distribution:

$$\mathbf{Y}|\mathbf{X} \sim \mathbf{NegBin}(\mathbf{n}\bar{\mathbf{x}} + \alpha, \mathbf{n} + \beta) \tag{16}$$

Bayes Estimator for Posterior Predictive:

$$\bar{\mathbf{Y}}|\mathbf{X} \sim \mathbf{NegBin}(\mathbf{n}\bar{\mathbf{x}} + \alpha, \mathbf{n} + \beta) \tag{17}$$

### 4.2.2. Precautionary Loss Function

In equations 18, 19, and 20 the mean, $\mu_{PLF,PG}$ (see 31), is the Bayes' estimator found by using the appropriate equation found in section 2.1.

Posterior Distribution:

$$\lambda|\mathbf{X} \sim Gamma(n\bar{x}+\alpha, n+\beta) \text{ where,}$$
$$\mu_{PLF,PG} = \frac{\sqrt{(n\bar{x}+\alpha)(n\bar{x}+\alpha)^2}}{n+\beta} \quad (18)$$

Posterior Predictive Distribution:

$$\mathbf{Y|X} \sim \mathbf{NegBin(n\bar{x}+\alpha, n+\beta)} \quad (19)$$

Bayes Estimator for Posterior Predictive:

$$\mathbf{\bar{Y}|X} \sim \mathbf{NegBin(n\bar{x}+\alpha, n+\beta)} \quad (20)$$

### 4.2.3. Linex Loss Function

In equations 21, 22, and 23 the mean, $\mu_{LLF,PG}$ (see 32), is the Bayes' estimator found by using the appropriate equation found in section 2.1.

Posterior Distribution:

$$\lambda|\mathbf{X} \sim Gamma(n\bar{x}+\alpha, n+\beta) \text{ where,}$$
$$\mu_{LLF,PG} = -\frac{1}{c}ln\left[\frac{\beta^{\alpha}}{(c+\beta)^{\alpha}}\right] \quad (21)$$

Posterior Predictive Distribution:

$$\mathbf{Y|X} \sim \mathbf{NegBin(n\bar{x}+\alpha, n+\beta)} \quad (22)$$

Bayes Estimator for Posterior Predictive:

$$\mathbf{\bar{Y}|X} \sim \mathbf{NegBin(n\bar{x}+\alpha, n+\beta)} \quad (23)$$

# 5. Simulations and Results

In this section sensitivity analyses of the hyper-parameters and sample size are ran for the Bayesian EWMA and Bayesian CUSUM control charts. These charts are designed to obtain $ARL_0 = 370$ and are assessed for different shift sizes ($\delta$ = 0 to 2.5 by 0.25). The hyper-parameter analysis considers the Poisson conjugate-based Bayesian CUSUM and EWMA control charts along with the Gaussian conjugate-based CUSUM control chart. For the simulations under each analysis, $m = 10,000$ iterations were ran to calculate the performance measurements of sample size $n = 10$.

## 5.1. Normal Conjugate

The Guassian conjugate-based CUSUM chart uses $h = 6$ to obtain the desired $ARL_0$. When generating the initial data we use a standard normal ($N(0,1)$), where the out-of-control mean is $\mu_1 = \mu + \delta\sigma$ and choices for the hyper-parameters are $\mu_0 = \{5, 10, 15\}$ and $\sigma = \{2, 4, 6\}$.

For the sample size analysis under both control charts we choose $n = \{5, 10, 20, 30\}$. Simulations for the EWMA chart calculates these measurements under varying smoothing parameter values as well ($\tau = 0.15, 0.30, 0.70$).

| Shifts | $\mu_0 = 5, \sigma_0 = 2$ | | | | $\mu_0 = 10, \sigma_0 = 4$ | | | | $\mu_0 = 15, \sigma_0 = 6$ | | | |
|---|---|---|---|---|---|---|---|---|---|---|---|---|
| | ARL | SDRL | ATS | SDTS | ARL | SDRL | ATS | SDTS | ARL | SDRL | ATS | SDTS |
| SELF | | | | | | | | | | | | |
| 0 | 382.56 | 314.006 | 2.10E-07 | 1.85E-06 | 377.585 | 310.835 | 2.06E-07 | 1.84E-06 | 381.314 | 311.04 | 1.96E-07 | 1.80E-06 |
| 0.25 | 25.1489 | 6.73527 | 2.18E-07 | 1.89E-06 | 25.112 | 6.68524 | 2.29E-07 | 1.98E-06 | 25.226 | 6.69327 | 1.67E-07 | 1.66E-06 |
| 0.5 | 12.2902 | 2.39349 | 1.98E-07 | 1.80E-06 | 12.3517 | 2.42792 | 2.09E-07 | 1.85E-06 | 12.2861 | 2.36657 | 1.97E-07 | 1.80E-06 |
| 0.75 | 7.9762 | 1.32674 | 1.79E-07 | 1.71E-06 | 7.9855 | 1.33375 | 2.10E-07 | 1.85E-06 | 7.9705 | 1.31796 | 2.13E-07 | 1.87E-06 |
| 1 | 5.8455 | 0.883872 | 2.29E-07 | 1.94E-06 | 5.8419 | 0.878012 | 2.12E-07 | 1.86E-06 | 5.8669 | 0.897321 | 2.18E-07 | 1.89E-06 |
| 1.25 | 4.5644 | 0.663365 | 1.90E-07 | 1.76E-06 | 4.5644 | 0.673092 | 1.94E-07 | 1.78E-06 | 4.5722 | 0.668421 | 2.18E-07 | 1.89E-06 |
| 1.5 | 3.7126 | 0.549 | 1.73E-07 | 1.68E-06 | 3.7255 | 0.555653 | 2.29E-07 | 3.63E-06 | 3.7143 | 0.549432 | 2.24E-07 | 1.92E-06 |
| 1.75 | 3.1044 | 0.411462 | 2.01E-07 | 1.81E-06 | 3.1065 | 0.412017 | 2.34E-07 | 1.96E-06 | 3.1091 | 0.405706 | 1.93E-07 | 1.78E-06 |
| 2 | 2.6992 | 0.468529 | 2.04E-07 | 1.83E-06 | 2.6991 | 0.469637 | 1.93E-07 | 1.78E-06 | 2.6859 | 0.473963 | 1.91E-07 | 1.77E-06 |
| 2.25 | 2.2077 | 0.405661 | 2.03E-07 | 1.82E-06 | 2.2124 | 0.409251 | 2.16E-07 | 1.97E-06 | 2.2192 | 0.413946 | 2.27E-07 | 1.93E-06 |
| 2.5 | 2.0136 | 0.143579 | 2.00E-07 | 1.81E-06 | 2.013 | 0.143635 | 2.15E-07 | 1.87E-06 | 2.0134 | 0.145672 | 1.96E-07 | 1.79E-06 |
| PLF | | | | | | | | | | | | |
| 0 | 386.091 | 317.271 | 1.73E-07 | 1.68E-06 | 383.807 | 308.288 | 2.36E-07 | 1.96E-06 | 382.062 | 309.744 | 2.19E-07 | 1.90E-06 |
| 0.25 | 25.0807 | 6.82654 | 1.90E-07 | 1.76E-06 | 25.2625 | 6.75559 | 2.30E-07 | 1.94E-06 | 25.3573 | 6.90383 | 2.06E-07 | 1.84E-06 |
| 0.5 | 12.293 | 2.43589 | 2.17E-07 | 1.88E-06 | 12.2928 | 2.40106 | 1.98E-07 | 1.80E-06 | 12.3003 | 2.41058 | 1.66E-07 | 1.66E-06 |
| 0.75 | 7.9805 | 1.33301 | 1.65E-07 | 1.65E-06 | 7.9887 | 1.33902 | 2.06E-07 | 1.88E-06 | 7.9786 | 1.31436 | 1.93E-07 | 1.78E-06 |
| 1 | 5.8491 | 0.897513 | 1.98E-07 | 1.80E-06 | 5.8279 | 0.899267 | 1.96E-07 | 1.80E-06 | 5.8588 | 0.884343 | 1.83E-07 | 1.73E-06 |
| 1.25 | 4.5609 | 0.6682 | 2.03E-07 | 1.83E-06 | 4.5705 | 0.667705 | 2.23E-07 | 1.91E-06 | 4.583 | 0.666417 | 1.65E-07 | 1.65E-06 |
| 1.5 | 3.7241 | 0.550617 | 2.23E-07 | 1.91E-06 | 3.7171 | 0.55106 | 1.90E-07 | 1.77E-06 | 3.7208 | 0.557716 | 1.81E-07 | 1.73E-06 |
| 1.75 | 3.1132 | 0.415194 | 1.95E-07 | 1.79E-06 | 3.1046 | 0.402068 | 1.89E-07 | 1.76E-06 | 3.1096 | 0.411081 | 2.07E-07 | 1.84E-06 |
| 2 | 2.6909 | 0.471123 | 2.23E-07 | 1.91E-06 | 2.7013 | 0.468058 | 2.26E-07 | 1.92E-06 | 2.6941 | 0.469601 | 1.80E-07 | 1.72E-06 |
| 2.25 | 2.2127 | 0.409706 | 1.83E-07 | 1.73E-06 | 2.2143 | 0.410336 | 1.96E-07 | 1.79E-06 | 2.2166 | 0.411928 | 1.92E-07 | 1.77E-06 |
| 2.5 | 2.0132 | 0.146375 | 2.00E-07 | 1.81E-06 | 2.0151 | 0.149238 | 1.99E-07 | 1.81E-06 | 2.016 | 0.144028 | 1.81E-07 | 1.72E-06 |
| LLF | | | | | | | | | | | | |
| 0 | 382.633 | 310.757 | 2.16E-07 | 1.88E-06 | 379.757 | 312.008 | 1.98E-07 | 1.80E-06 | 381.621 | 311.286 | 2.05E-07 | 1.84E-06 |
| 0.25 | 25.2814 | 6.8093 | 1.90E-07 | 1.77E-06 | 25.1894 | 6.71061 | 1.55E-07 | 1.60E-06 | 25.222 | 6.70407 | 1.66E-07 | 1.65E-06 |
| 0.5 | 12.2979 | 2.42684 | 2.14E-07 | 1.87E-06 | 12.2589 | 2.39922 | 1.81E-07 | 1.72E-06 | 12.2623 | 2.35586 | 1.86E-07 | 1.75E-06 |
| 0.75 | 7.9646 | 1.31482 | 2.00E-07 | 1.81E-06 | 7.9809 | 1.32058 | 1.82E-07 | 1.73E-06 | 7.9879 | 1.32399 | 2.06E-07 | 1.84E-06 |
| 1 | 5.8316 | 0.88828 | 1.88E-07 | 1.76E-06 | 5.8337 | 0.889294 | 1.83E-07 | 1.73E-06 | 5.8465 | 0.890471 | 2.02E-07 | 1.82E-06 |
| 1.25 | 4.5581 | 0.669645 | 2.14E-07 | 1.87E-06 | 4.5811 | 0.660926 | 1.46E-07 | 1.55E-06 | 4.5675 | 0.671896 | 1.85E-07 | 1.74E-06 |
| 1.5 | 3.723 | 0.555041 | 2.04E-07 | 1.83E-06 | 3.7261 | 0.549253 | 1.74E-07 | 1.69E-06 | 3.7283 | 0.551615 | 1.88E-07 | 1.76E-06 |
| 1.75 | 3.1041 | 0.404306 | 2.15E-07 | 1.88E-06 | 3.108 | 0.406615 | 1.96E-07 | 1.80E-06 | 3.1026 | 0.41361 | 1.81E-07 | 1.72E-06 |
| 2 | 2.6947 | 0.470204 | 2.07E-07 | 1.84E-06 | 2.6936 | 0.472355 | 1.81E-07 | 1.72E-06 | 2.6927 | 0.472934 | 2.34E-07 | 1.96E-06 |
| 2.25 | 2.2067 | 0.405679 | 2.04E-07 | 1.83E-06 | 2.2029 | 0.402159 | 1.41E-07 | 1.53E-06 | 2.2195 | 0.414391 | 1.86E-07 | 1.75E-06 |
| 2.5 | 2.0152 | 0.144114 | 2.43E-07 | 2.01E-06 | 2.0132 | 0.136476 | 1.64E-07 | 1.64E-06 | 2.016 | 0.144028 | 2.08E-07 | 1.85E-06 |

Table 4: ARL, SDRL, ATS, and SDTS Values for Bayesian CUSUM Chart Hyper-Parameter Sensitivity Analysis with Normal Conjugate

Table 4 shows simulation results from the Bayesian CUSUM control chart under a normal conjugate prior. A constant value of $h = 6$ was used to obtain in-control ARL results of around 370. The prior mean and variance were adjusted for sensitivity analysis of the hyper-parameters while applying shifts to the in-control mean ($\mu_0$). For all the loss functions, the initial shift in the mean returns a drastic drop in the ARL and SDRL. After the first shift, as the shifts increase the ARL and SDRL gradually decrease. This shows that a relatively small shift is detectable when using the Bayesian CUSUM control chart and detection is consistent over several loss functions. Results also show that as you adjust the input for the hyper-parameters, this control chart per-forms consistently. ATS and SDTS values, in seconds, are not significantly different within each loss function, but overall the Linex loss function has values that are noticeably smaller than the squared error and precautionary loss functions.

| | | | | | | | Shifts | | | | | |
|---|---|---|---|---|---|---|---|---|---|---|---|---|
| n | h | 0 | 0.25 | 0.5 | 0.75 | 1 | 1.25 | 1.5 | 1.75 | 2 | 2.25 | 2.5 |
| | | | | | | | SELF | | | | | |
| 5 | 8.4 | 375.399 | 37.1613 | 18.255 | 11.9493 | 8.8191 | 6.9268 | 5.6775 | 4.784 | 4.1443 | 3.6073 | 3.1807 |
| | | (308.228) | (11.9545) | (4.21535) | (2.3375) | (1.54802) | (1.11177) | (0.867118) | (0.717038) | (0.599731) | (0.551984) | (0.44525) |
| | | 2.07E-07 | 2.18E-07 | 1.76E-07 | 2.06E-07 | 2.01E-07 | 2.46E-07 | 1.69E-07 | 2.16E-07 | 2.11E-07 | 2.11E-07 | 1.80E-07 |
| | | (1.86E-06) | (1.89E-06) | (1.70E-06) | (1.84E-06) | (1.81E-06) | (2.01E-06) | (1.67E-06) | (1.88E-06) | (1.86E-06) | (1.87E-06) | (1.72E-06) |
| 10 | 6 | 381.085 | 25.2921 | 12.3169 | 8.002 | 5.8504 | 4.583 | 3.7143 | 3.1071 | 2.6879 | 2.2137 | 2.015 |
| | | (313.336) | (6.77449) | (2.4378) | (1.34781) | (0.885223) | (0.662202) | (0.547244) | (0.410889) | (0.47444) | (0.410161) | (0.140624) |
| | | 2.60E-07 | 1.67E-07 | 2.19E-07 | 1.97E-07 | 1.75E-07 | 1.89E-07 | 1.82E-07 | 1.97E-07 | 1.78E-07 | 1.78E-07 | 1.72E-07 |
| | | (2.06E-06) | (1.66E-06) | (1.89E-06) | (1.80E-06) | (1.69E-06) | (1.76E-06) | (1.73E-06) | (1.80E-06) | (1.71E-06) | (1.71E-06) | (1.68E-06) |
| 20 | 4.18 | 369.377 | 16.9652 | 8.1781 | 5.2669 | 3.809 | 2.9699 | 2.2954 | 2.0001 | 1.8121 | 1.2501 | 1.0119 |
| | | (299.4) | (3.82957) | (1.40384) | (0.79364) | (0.562245) | (0.379465) | (0.456441) | (0.126095) | (0.390632) | (0.43307) | (0.108436) |
| | | 1.90E-07 | 1.57E-07 | 2.11E-07 | 2.14E-07 | 1.92E-07 | 2.07E-07 | 1.93E-07 | 2.03E-07 | 1.50E-07 | 1.91E-07 | 2.09E-07 |
| | | (1.76E-06) | (1.60E-06) | (1.86E-06) | (1.88E-06) | (1.77E-06) | (1.84E-06) | (1.78E-06) | (1.83E-06) | (1.57E-06) | (1.77E-06) | (1.85E-06) |
| 30 | 3.4 | 373.031 | 13.5429 | 6.4694 | 4.1425 | 2.9955 | 2.1861 | 1.9625 | 1.4297 | 1.0215 | 1 | 1 |
| | | (304.37) | (2.84098) | (1.02921) | (0.60116) | (0.386885) | (0.389187) | (0.193633) | (0.495033) | (0.145044) | (0) | (0) |
| | | 2.33E-07 | 2.25E-07 | 2.08E-07 | 2.14E-07 | 2.09E-07 | 2.31E-07 | 2.59E-07 | 1.97E-07 | 2.14E-07 | 2.70E-07 | 2.24E-07 |
| | | (1.95E-06) | (1.92E-06) | (1.84E-06) | (1.87E-06) | (1.85E-06) | (1.95E-06) | (2.06E-06) | (1.79E-06) | (1.87E-06) | (2.10E-06) | (1.91E-06) |
| | | | | | | | PLF | | | | | |
| 5 | 8.4 | 375.604 | 37.207 | 18.2401 | 11.9825 | 8.7833 | 6.9291 | 5.6663 | 4.7933 | 4.1237 | 3.6066 | 3.1778 |
| | | (310.609) | (12.1227) | (4.24745) | (2.34252) | (1.53191) | (1.10873) | (0.860781) | (0.703829) | (0.595649) | (0.554289) | (0.440213) |
| | | 2.09E-07 | 1.94E-07 | 2.10E-07 | 2.01E-07 | 1.70E-07 | 1.64E-07 | 1.56E-07 | 1.96E-07 | 1.87E-07 | 1.83E-07 | 2.08E-07 |
| | | (1.85E-06) | (1.79E-06) | (1.85E-06) | (1.82E-06) | (1.67E-06) | (1.64E-06) | (1.60E-06) | (1.79E-06) | (1.75E-06) | (1.73E-06) | (1.85E-06) |
| 10 | 6 | 377.371 | 25.1675 | 12.2698 | 7.9768 | 5.8525 | 4.5655 | 3.7289 | 3.1078 | 2.6988 | 2.2135 | 2.0124 |
| | | (303.993) | (6.62654) | (2.41851) | (1.33726) | (0.892381) | (0.662955) | (0.557499) | (0.41398) | (0.469338) | (0.410022) | (0.138008) |
| | | 1.88E-07 | 2.17E-07 | 2.01E-07 | 2.05E-07 | 1.87E-07 | 2.10E-07 | 1.84E-07 | 2.08E-07 | 1.65E-07 | 1.84E-07 | 1.93E-07 |
| | | (1.75E-06) | (1.90E-06) | (1.82E-06) | (1.83E-06) | (1.75E-06) | (1.85E-06) | (1.74E-06) | (1.85E-06) | (1.65E-06) | (1.74E-06) | (1.78E-06) |
| 20 | 4.18 | 368.62 | 16.9351 | 8.1502 | 5.2623 | 3.8054 | 2.9668 | 2.2932 | 1.9982 | 1.8066 | 1.2553 | 1.0122 |
| | | (297.546) | (3.80162) | (1.39422) | (0.797307) | (0.566684) | (0.388713) | (0.455449) | (0.131136) | (0.394964) | (0.43603) | (0.109778) |
| | | 1.88E-07 | 1.96E-07 | 2.04E-07 | 1.98E-07 | 3.03E-07 | 1.76E-07 | 2.16E-07 | 2.10E-07 | 1.97E-07 | 2.08E-07 | 1.71E-07 |
| | | (1.76E-06) | (1.79E-06) | (1.83E-06) | (1.80E-06) | (2.22E-06) | (1.70E-06) | (1.88E-06) | (1.86E-06) | (1.80E-06) | (1.85E-06) | (1.68E-06) |
| 30 | 3.41 | 367.25 | 13.5895 | 6.5236 | 4.1415 | 3.0024 | 2.185 | 1.9616 | 1.447 | 1.0208 | 1.0001 | 1 |
| | | (295.354) | (2.83235) | (1.0291) | (0.603389) | (0.3914) | (0.389326) | (0.19423) | (0.497183) | (0.142714) | (0.0099995) | (0) |
| | | 1.80E-07 | 2.03E-07 | 1.80E-07 | 2.09E-07 | 2.17E-07 | 2.38E-07 | 1.98E-07 | 1.91E-07 | 2.22E-07 | 1.98E-07 | 1.88E-07 |
| | | (1.72E-06) | (1.82E-06) | (1.72E-06) | (1.85E-06) | (1.89E-06) | (1.97E-06) | (1.80E-06) | (1.77E-06) | (1.90E-06) | (1.80E-06) | (1.76E-06) |
| | | | | | | | LLF | | | | | |
| 5 | 8.4 | 376.357 | 37.1803 | 18.2026 | 11.9679 | 8.8241 | 6.9209 | 5.6815 | 4.8087 | 4.1361 | 3.6015 | 3.1848 |
| | | (314.896) | (12.0124) | (4.27094) | (2.32152) | (1.53217) | (1.10256) | (0.868595) | (0.721183) | (0.592095) | (0.547264) | (0.444352) |
| | | 1.52E-07 | 1.56E-07 | 2.10E-07 | 2.15E-07 | 1.97E-07 | 1.75E-07 | 1.75E-07 | 1.93E-07 | 1.75E-07 | 1.71E-07 | 1.53E-07 |
| | | (1.58E-06) | (1.61E-06) | (1.86E-06) | (1.88E-06) | (1.80E-06) | (1.69E-06) | (1.69E-06) | (1.78E-06) | (1.69E-06) | (1.68E-06) | (1.59E-06) |
| 10 | 6 | 378.459 | 25.3499 | 12.2567 | 7.9641 | 5.8521 | 4.5691 | 3.7212 | 3.1041 | 2.6955 | 2.2024 | 2.0124 |
| | | (312.249) | (6.70681) | (2.37525) | (1.33799) | (0.884435) | (0.670392) | (0.551245) | (0.409955) | (0.469446) | (0.402038) | (0.129793) |
| | | 1.76E-07 | 1.60E-07 | 2.19E-07 | 1.86E-07 | 1.81E-07 | 1.81E-07 | 1.89E-07 | 1.84E-07 | 2.19E-07 | 1.90E-07 | 1.75E-07 |
| | | (1.70E-06) | (1.62E-06) | (1.90E-06) | (1.75E-06) | (1.73E-06) | (1.72E-06) | (1.76E-06) | (1.74E-06) | (1.89E-06) | (1.77E-06) | (1.70E-06) |
| 20 | 4.18 | 371.933 | 16.9739 | 8.1679 | 5.2633 | 3.8142 | 2.9723 | 2.2925 | 2.0014 | 1.8113 | 1.2497 | 1.014 |
| | | (310.142) | (3.86676) | (1.3851) | (0.791564) | (0.560605) | (0.382011) | (0.45513) | (0.116611) | (0.39127) | (0.432839) | (0.11749) |
| | | 1.80E-07 | 2.58E-07 | 1.53E-07 | 1.83E-07 | 1.98E-07 | 1.94E-07 | 2.01E-07 | 1.69E-07 | 2.04E-07 | 1.85E-07 | 1.90E-07 |
| | | (1.72E-06) | (2.05E-06) | (1.58E-06) | (1.73E-06) | (1.80E-06) | (1.79E-06) | (1.82E-06) | (1.67E-06) | (1.83E-06) | (1.74E-06) | (1.76E-06) |
| 30 | 3.41 | 368.33 | 13.5945 | 6.5128 | 4.1572 | 3.0067 | 2.1906 | 1.967 | 1.4585 | 1.0224 | 1.0001 | 0.9999 |
| | | (304.737) | (2.84954) | (1.02442) | (0.604722) | (0.382172) | (0.392774) | (0.182513) | (0.498275) | (0.147981) | (0.0099995) | (0.0099995) |
| | | 1.88E-07 | 1.68E-07 | 2.27E-07 | 1.63E-07 | 1.71E-07 | 1.58E-07 | 1.92E-07 | 1.70E-07 | 1.65E-07 | 1.69E-07 | 2.31E-07 |
| | | (1.76E-06) | (1.67E-06) | (1.92E-06) | (1.64E-06) | (1.68E-06) | (1.61E-06) | (1.77E-06) | (1.67E-06) | (1.65E-06) | (1.67E-06) | (1.95E-06) |

Table 5: ARL, SDRL, ATS, and SDTS Values for Bayesian CUSUM Chart Sample Size Sensitivity Analysis with Normal Conjugate

Similar trends about the Bayesian CUSUM chart that were noticed in table 4 can also be seen in table 5. By decreasing the $h$ value as the sample size increases, we obtain $ARL_0$ values around 370. This is intuitive since increasing sample size typically delays detection, thus shrinking the bounds for the control limits allows for timely detection. As before, we notice that there is an immediate drop in ARL when the initial mean shift is applied and a gradual decrease after for all loss functions. We notice that as we increase $n$ while applying the mean shifts, the drop in the ARL is larger in the larger samples. This shows that detection ability grows with sample size. ATS and SDTS are also effected by increasing sample size; detection time decreases as sample size increases.

## *5.2. Poisson Conjugate*

Our next simulations will be ran for the Bayesian control charts under the Poisson-Gamma case. We will design both a Bayesian EWMA and Bayesian CUSUM chart to obtain a defined in-control ARL and assess their performances for the different shift sizes. We use values of $\mu_0$ and $\sigma_0$ to solve for our values of $\alpha$ and $\beta$, these values will be used as our prior information. These values are $\alpha = [4, 16, 36]$ and $\beta = \left[\frac{5}{4}, \frac{5}{8}, \frac{5}{12}\right]$ as calculated in section 6.1.2. Our initial data will be generated from a Poisson distribution with parameter $\lambda$ set as constant.

The sample size analysis will use the same values for $n$ as in section 5.1 and the values for $\tau$ and $h$ will be determined based on our desired $ARL_0$.

# 6. Appendix

## 6.1. Poisson-Gamma Derivations

Poisson Likelihood:
$$x|\lambda \sim Poisson(\lambda)$$
$$f(x|\lambda) = \frac{e^{-n\lambda}\ \lambda^{\sum_{i=1}^{n} x_i}}{\prod_{i=1}^{n} x_i!} \tag{24}$$

Gamma Prior:
$$\lambda \sim Gamma(\alpha, \beta)$$
$$f(\lambda) = \frac{\beta^\alpha}{\Gamma(\alpha)} \lambda^{\alpha-1} e^{-\beta\lambda} \tag{25}$$

Gamma Posterior:
$$\lambda|x \sim Gamma(n\bar{x} + \alpha, n + \beta)$$
$$f(\lambda|x) = \frac{\beta^\alpha}{\Gamma(\alpha)} \lambda^{(n\bar{x}+\alpha)-1} e^{-(n+\beta)\lambda} \tag{26}$$

Posterior Predictive for Poisson Conjugate:

$$y|x \sim Poisson(\lambda)$$
$$f(y|x) = \frac{e^{-\lambda}\ \lambda^y}{y!} \tag{27}$$

$$f(y|x) = \int f(\lambda|x) f(y|\lambda) d\lambda$$

$$= \int \frac{(\beta+n)^{n\bar{x}+\alpha}}{\Gamma(n\bar{x}+\alpha)} e^{-(n+\beta)\lambda} \lambda^{(n\bar{x}+\alpha)-1} \frac{e^{-\lambda} \lambda^y}{y!} d\lambda$$

$$= \frac{(\beta+n)^{n\bar{x}+\alpha}}{\Gamma(n\bar{x}+\alpha) y!} \int e^{-(n+\beta)} e^{-\lambda} \lambda^{(n\bar{x}+\alpha)-1} \lambda^y d\lambda$$

$$\text{let c} = \frac{(\beta+n)^{n\bar{x}+\alpha}}{\Gamma(n\bar{x}+\alpha) y!}, \quad c \int e^{-(n+\beta)\lambda - \lambda} \lambda^{(n\bar{x}+\alpha)-1+y} d\lambda$$

$$= c \int e^{-(n+\beta+1)\lambda} \lambda^{(n\bar{x}+y+\alpha)-1} d\lambda \qquad (28)$$

$$= \frac{(\beta+n)^{n\bar{x}+\alpha}}{\Gamma(n\bar{x}+\alpha) y!} \cdot \frac{\Gamma(n\bar{x}+\alpha)+y}{(\beta+n+1)^{n\bar{x}+\alpha+y}}$$

$$= \frac{\Gamma(n\bar{x}+\alpha+y)}{\Gamma(n\bar{x}+\alpha) y!} \cdot \frac{(\beta+n)^{n\bar{x}+\alpha}}{(\beta+n+1)^{n\bar{x}+\alpha+y}}$$

$$= \frac{(n\bar{x}+\alpha+y)!}{(n\bar{x}+\alpha)! y!} \cdot \left(\frac{\beta+n}{\beta+n+1}\right)^{n\bar{x}+\alpha} \left(\frac{1}{\beta+n+1}\right)^y$$

$$\binom{n\bar{x}+\alpha+y-1}{y-1} \cdot \left(1 - \frac{1}{\beta+n+1}\right)^{n\bar{x}+\alpha} \left(\frac{1}{\beta+n+1}\right)^y$$

$$f(y|x) = \binom{n\bar{x}+\alpha+y-1}{y-1} \left(1 - \frac{1}{\beta+n+1}\right)^{n\bar{x}+\alpha} \left(\frac{1}{\beta+n+1}\right)^y \qquad (29)$$

$$y|x \sim NegBin(n\bar{x}+\alpha, n+\beta)$$

*6.1.1. Loss Functions Best Estimators*

SELF:

$$\hat{\lambda}^* = E[\lambda|x] = \frac{n\bar{x}+\alpha}{n+\beta},$$

$$\left[\mu_{SELF,PG} = \hat{\lambda}^* = \frac{n\bar{x}+\alpha}{n+\beta}\right] \qquad (30)$$

PLF:

$$\hat{\lambda}^* = \sqrt{E[\lambda^2|x]}$$

$$VAR[\lambda|x] = E[\lambda^2|x] - E[\lambda|x]^2$$

$$\frac{n\bar{x}+\alpha}{(n+\beta)^2} = E[\lambda^2|x] - \left(\frac{n\bar{x}+\alpha}{n+\beta}\right)^2$$

$$\frac{n\bar{x}+\alpha}{(n+\beta)^2} = E[\lambda^2|x] - \frac{(n\bar{x}+\alpha)^2}{(n+\beta)^2} \quad (31)$$

$$E[\lambda^2|x] = \frac{(n\bar{x}+\alpha) + (n\bar{x}+\alpha)^2}{(n+\beta)^2}$$

$$\sqrt{E[\lambda^2|x]} = \sqrt{\frac{(n\bar{x}+\alpha) + (n\bar{x}+\alpha)^2}{(n+\beta)^2}}$$

$$\left[\mu_{PLF,PG} = \hat{\lambda}^* = \frac{\sqrt{(n\bar{x}+\alpha) + (n\bar{x}+\alpha)^2}}{(n+\beta)}\right]$$

LLF:

$$\hat{\lambda}^* = -\frac{1}{c}ln\left[E[e^{-c\lambda}]\right]$$

$$E[e^{-c\lambda}] = \int e^{-c\lambda} \cdot \frac{\beta^\alpha}{\Gamma(\alpha)} \lambda^{\alpha-1} e^{-\beta\lambda} d\lambda$$

$$= \frac{\beta^\alpha}{\Gamma(\alpha)} \int e^{-c\lambda} \lambda^{\alpha-1} e^{-\beta\lambda} d\lambda$$

$$= \frac{\beta^\alpha}{\Gamma(\alpha)} \int e^{-(c+\beta)\lambda} \lambda^{\alpha-1} d\lambda \quad (32)$$

$$= \frac{\beta^\alpha}{\cancel{\Gamma(\alpha)}} \cdot \frac{\cancel{\Gamma(\alpha)}}{(c+\beta)^\alpha}$$

$$\left[\mu_{LLF,PG} = \hat{\lambda}^* = -\frac{1}{c}ln\left[\frac{\beta^\alpha}{(c+\beta)^\alpha}\right]\right]$$

### 6.1.2. Solving for α and β

Given $\mu_0 = [5, 10, 15]$ and $\sigma_0^2 = [4, 16, 36]$:

$\underline{\mu_0 = 5,\ \sigma_0^2 = 4}$

$$5 = \frac{\alpha}{\beta} \Rightarrow \alpha = 5\beta$$
$$4 = \frac{\alpha}{\beta^2} \Rightarrow 4 = \frac{5\beta}{\beta^2} \Rightarrow \left[\beta = \frac{5}{4}\right] \tag{33}$$
$$5 = \frac{\alpha}{5/4} \Rightarrow [\alpha = 4]$$

$\underline{\mu_0 = 10,\ \sigma_0^2 = 16}$

$$10 = \frac{\alpha}{\beta} \Rightarrow \alpha = 10\beta$$
$$16 = \frac{\alpha}{\beta^2} \Rightarrow 16 = \frac{10\beta}{\beta^2} \Rightarrow \left[\beta = \frac{5}{8}\right] \tag{34}$$
$$10 = \frac{\alpha}{5/8} \Rightarrow [\alpha = 16]$$

$\underline{\mu_0 = 15,\ \sigma_0^2 = 36}$

$$15 = \frac{\alpha}{\beta} \Rightarrow \alpha = 15\beta$$
$$36 = \frac{\alpha}{\beta^2} \Rightarrow 36 = \frac{15\beta}{\beta^2} \Rightarrow \left[\beta = \frac{5}{12}\right] \tag{35}$$
$$15 = \frac{\alpha}{5/12} \Rightarrow [\alpha = 36]$$